\documentstyle{article}
\DeclareSymbolFont{EUrm}{U}{eur}{m}{n}
\DeclareMathSymbol{\deltaF}{\mathchar}{EUrm}{"0E}
\DeclareMathSymbol{\thetaF}{\mathchar}{EUrm}{"12}
\begin{document}

\newcommand{\rhob}{\mbox{\boldmath$\rho$}}
\newcommand{\alphab}{\hat{\alpha}}
\newcommand{\betab}{\hat{\beta}}
\newcommand{\betac}{\hat{\cal B}}
\newcommand{\etab}{\hat{\eta}}
\newcommand{\gammab}{\hat{\gamma}}
\newcommand{\G}{{\bf G}}
\newcommand{\bk}{{\bf k}}
\newcommand{\bq}{{\bf q}}
\newcommand{\bA}{{\bf A}}
\newcommand{\bB}{{\bf B}}
\newcommand{\cE}{{\cal E}}
\newcommand{\Gd}{\tilde{\bf G}}
\newcommand{\g}{{\bf G}^0}
\newcommand{\Q}{{\bf Q}}
\newcommand{\Gr}{{\bf G}_1}
\newcommand{\Grd}{\tilde{\bf G}_1}
\newcommand{\Gs}{{\bf G}_2}
\newcommand{\Gsd}{\tilde{\bf G}_2}
\newcommand{\Gsa}{{\bf G}_{2a}}
\newcommand{\Gsad}{\tilde{\bf G}_{2a}}
\newcommand{\Gsb}{{\bf G}_{2b}}
\newcommand{\Gsbd}{\tilde{\bf G}_{2b}}
\newcommand{\ds}
{
\overline{
\overline
{(\Delta\sigma)^2}
}
}
\newcommand{\de}{{\mathrm d}}
\newcommand{\ex}{{\mathrm e}}
\begin{titlepage}

\title{ Magnetic Properties of Dilute Alloys: Equations for
Magnetization and its Structural Fluctuations}
\author{I.~Vakarchuk$^1$, V.~Tkachuk$^2$, T.~Kuliy$^3$}
\maketitle\thispagestyle{empty}
\begin{small}
\begin{center}
Department for Theoretical Physics, Ivan Franko L'viv State University
\\ 12 Drahomanov Str., L'viv UA--290005, Ukraine
\\ phone: +380-322-728080, fax: +380-322-727981
\\ $^1$ e--mail: Chair@KTF.Franko.Lviv.UA, $^2$ e--mail: Tkachuk@KTF.Franko.Lviv.UA,
\\ $^3$ e--mail: Kuliy@KTF.Franko.Lviv.UA
\\[1.5ex] PACS numbers: 75.10.JM, 75.30.Ds
\end{center}
\end{small}
\begin{abstract}
The dilute Heisenberg ferromagnet is studied taking into account fluctuations
of magnetization caused by disorder. A self--consistent system of equations for
magnetization and its mean quadratic fluctuations is derived within the
configurationally averaged two--time temperature Green's function method.
This system of equations is analised at low concentration of non--magnetic 
impurities. Mean relative quadratic fluctuations of magnetization are revealed 
to be proportional to the square of concentration of impurities.
\end{abstract}\end{titlepage}

\renewcommand{\theequation}{\arabic{equation}}
\section*{Introduction}

The method of configurationally averaged Green's function developed by
Kaneyoshi \cite{Kan69.41, Kan69.42} to study a dilute ferromagnet was used
successfully by several authors \cite{Kan72, Kan83, Vak-Tk90, JMMM95} to
describe disordered magnetic systems. One of the most frequently used approximations
within the frames of this method is neglecting the fluctuations of
magnetization caused by disorder \cite{Marg87, Marg88}. Note that at zero
temperature the magnetization for Heisenberg model tends to saturation and
fluctuations are suppressed.

In paper \cite{Kan83} it was shown that taking into account the
structural fluctuations of magnetization is necessary in order to explain the
anomalous behaviour of spin--wave stiffness constant in
amorphous ferromagnets. It was assumed that the structural
fluctuations of magnetization at each site are statistically independent and
are governed by a Gauss distribution function. In order to
obtain quantitative results one needs to take into account the
connection of the structural fluctuations of magnetization with the
fluctuations of structure. In our previous work \cite{JMMM95} this relation
was obtained within the framework of mean--field approximation. It allowed one
to analyse quantitatively the spin--wave spectrum of amorphous ferromagnets.

In the present paper we estimate the dependence of structural fluctuations of
magnetization on the concentration of non--magnetic sites in
a dilute ferromagnet. We suggest
a self--consistent set of equations for magnetization and its quadratic
fluctuations derived within the framework of the configurationally averaged 
two--time temperature Green's function method.

\renewcommand{\theequation}{\arabic{section}.\arabic{equation}}
\setcounter{equation}{0}
\section{Configurationally Averaged Green's Functions}

Let us consider a structurally disordered system of $N$ atoms in the volume
$V$ which is described by the isotropic Heisenberg Hamiltonian

\begin{equation} \label{Ham}
    H = -{1 \over 2} \sum_{i,j} J_{ij} n_i n_j{\bf S}_i{\bf S}_j
    -h \sum_{j} n_j S^z_j,
\end{equation}
where
$J_{ij}=J(|{\bf R}_i-{\bf R}_j|)$
is the exchange integral describing the interaction
between the $i$th and $j$th atoms, ${\bf S}_i$ is the spin operator
of the $i$th atom, $h$ is the external magnetic field, $n_i$ is a random
variable taking the value of 1 or 0 according to whether or not the site $i$
is occupied by a magnetic atom.

The Fourier--component of retarding two--time temperature Green's function
\begin{equation}\label{Green-l}
\ll l|l'\gg_t\equiv\ll n_l S^+_l|n_{l'} S^-_{l'}\gg_t=-i\thetaF(t)
\langle\left[S^+_l(t),S^-_{l'}(0)\right]\rangle n_l n_{l'}
\end{equation}
satisfies within the Tyablikov approximation the equation of motion

\begin{equation}  \label{G_ll'-eq}
(E-h)\ll l|l'\gg_E=2\deltaF_{l,l'}x_l +\sum_{j(\not=l)} J_{lj}
\left(x_j \ll l|l'\gg_E- x_l \ll j|l'\gg_E \right),
\end{equation}
where $x_i=<S^z_i>n_i$ is the magnetization of the $i$th site and
$x=\overline{x_i}$ is the mean magnetization. The overbar means the averaging
over configurations but due to self--averaging of magnetization we can
write
$$x={1\over N}\sum_i{x_i}={1\over N}\sum_i{<S^z_i>n_i}
={N_m\over N}{1\over N_m}\sum_{i(n_i=1)}{<S^z_i>}
=n\overline{\overline{<S^z_i>}}=n\sigma,$$
where double overline means the averaging over the magnetic sites only.
Extracting fluctuations
$\xi_i=x_i-x$ and coming to a momentum space
$${\ll\bq |\bq'\gg}=
{1\over N} \sum\limits_{i,j=1}^N
\ex^{-i \left(\bq{\bf R}_i-\bq'{\bf R}_j\right)}\ll i|j\gg,\ \ \
\xi_\bq = {1\over\sqrt{N}}\sum\limits_{j=1}^N
\xi_j \ex^{-i\bq{\bf R}_j},$$
we get
\begin{eqnarray}  \label{Green-k}
(E-E_0(\bq))\ll\bq |\bq'\gg
=2 x\deltaF(\bq -\bq') +
{2\over\sqrt{N}} \xi_{\bq-\bq'}+
 {1\over \sqrt{N}}\sum_\bk A(\bq,\bk)
\xi_{\bq-\bk}\ll\bk |\bq'\gg ,
\end{eqnarray}
where
$J(\bq )=\sum\limits_{\bf R} J(R) \ex^{-i \bq{\bf R}}$
is a Fourier transform of exchange integral,
$E_0(\bq )=h+xA(\bq,\bq )$
is the spin--wave spectrum of a non--dilute crystal and the notation
$A(\bq,\bk )=J(\bq -\bk )-J(\bk )$
is introduced for
convenience. All sums in equation (\ref{Green-k}) and hereafter in the reciprocal space
are taken over the Brillouin zone.
It is worth while remembering that there is just a lattice and
therefore all Fourier transformations from the real space to momentum one are
reversible.

Note that $x_i$ or $\xi_i$ include fluctuations of structure $n_i$
as well as fluctuations of magnetization $<S^z_i>$.
It is convenient to rewrite equation (\ref{Green-k}) in the matrix form:
\begin{equation}\label{G}
\G=\g+\g\rhob+\g\Q\G,
\end{equation}
where %\linebreak\\

$G_{\bf q,q'}=\ll\bq |{\bf q'}\gg$ is the unaveraged Green's function,\\

%\linebreak
$G^0_{\bf q,q'}=
{
\displaystyle{2x\deltaF}_{\bf q,q'}
\over
\displaystyle{E-E_0(\bf{q})}
}$
is the zero--approximation Green's function,\\

$\rho_{\bf q,q'}=
{
\displaystyle{\xi}_{\bf q-q'}
\over
\displaystyle{\sqrt{N}x}
}$
is the matrix of relative fluctuations of magnetization
and the matrix\\

$Q_{\bf q,q'}=
{
\displaystyle {A({\bf q,q'})}
\over
\displaystyle{2}
}
\rho_{\bf q,q'}$ is
proportional to relative fluctuations of magnetization and depends on the 
exchange interaction between spins. The zero--approximation Green's function 
of the dilute system has the same form as in the non--dilute case with 
the only difference that the mean magnetization  $x=n\sigma$ must be substituted in place 
of $\sigma$.
Averaging equation (\ref{G}) over configurations we obtain
\begin{equation}\label{G-a0}
\overline{\G}=\g+\g\overline{\Q\G}.
\end{equation}
To obtain an equation for the new unknown Green's function $\overline{\Q\G}$ we 
need to multiply equation (\ref{G}) by the matrix $\g\Q$ and 
to average the product over configurations. Substituting the result in 
expression (\ref{G-a0}) we get another expression for
$\overline{\G}$:
\begin{equation}\label{G-a1}
\overline{\G}=\g+\g\overline{\Q\g\rhob}+\g\overline{\Q\g\Q\G},
\end{equation}
which contains the matrix $\overline{\Q\g\Q\G}$. We can obtain
the equation for this averaged product in the same way and continuing this
iteration procedure we can get similarly as it is described in \cite{Kan69.42}
the expression for the configurationally averaged Green's function
\begin{equation}\label{G-a}
\overline{\G}=\left(1-\g\sum_{i=1}^\infty\overline{{\Delta\Q_i\Q}}
\right)^{-1}\g\left(1+\sum_{i=1}^\infty\overline{{\Delta\Q_i\rhob}}
\right),
\end{equation}
where
$$
\Delta\Q_i=\Q_i-\overline{\Q_i}, \ \ \ \Q_1=\Q\g, \ \ \
\Q_i=\Delta\Q_{i-1}\Q\g \ {\rm for} \ i>1.
$$
All three factors in equation (\ref{G-a}) are diagonal matrices in the
momentum space and we can rewrite this expression as
\begin{equation}\label{G_q-a}
[\overline{\G}]_\bq =
{[\g]_\bq 
\left(1+\sum\limits_{i=1}^\infty[\overline{\Delta\Q_i\rhob}]_\bq 
\right)
\over
1-[\g]_\bq\sum\limits_{i=1}^\infty[\overline{\Delta\Q_i\Q}]_\bq 
},
\end{equation}
where ${[\ldots]}_\bq\equiv{[\ldots]}_{\bf q,q}$ is a diagonal element of
the matrix.

Let us consider other Green's function $\Gr=\G\rhob$ which we will use further. To
obtain the equation for this function let us multiply the equation (\ref{G})
by $\rhob$. We get
\begin{equation}\label{Gr}
\Gr=\g(\rhob+\rhob^2)+\g\Q\Gr.
\end{equation}
An averaged value of $\Gr$ obtained within the similar iteration
procedure has the following form
\begin{equation}\label{Gr-a}
\overline{\Gr}=\left(1-\g\sum_{i=1}^\infty\overline{{\Delta\Q_i\Q}}
\right)^{-1}\g\left(\overline{\rhob^2}+
\sum_{i=1}^\infty\overline{{\Delta\Q_i(\rhob+\rhob^2)}}
\right)
\end{equation}
or
\begin{equation}\label{Gr_q-a}
[\overline{\Gr}]_\bq =
{
[\g]_\bq\left([\overline{\rhob^2}]_\bq +
\sum\limits_{i=1}^\infty[\overline{\Delta\Q_i(\rhob+\rhob^2)}]_\bq 
\right)
\over
1-[\g]_\bq\sum\limits_{i=1}^\infty[\overline{\Delta\Q_i\Q}]_\bq 
}
\end{equation}
for the diagonal elements in the momentum space.

Note that the matrix $\overline{\rhob^2}$ is proportional to the unit matrix:
\begin{equation}\label{r2-a}
[\overline{\rhob^2}]_\bq ={1\over N}\sum_\bk 
{\overline{\xi_{\bf q-k}\xi_{\bf k-q}}\over
x^2}={1\over N}\sum_{l}
{\overline{\xi_l^2}\over
x^2}={\overline{\xi^2}\over x^2}.
\end{equation}

\setcounter{equation}{0}
\section{Equations for Magnetization and its Structural Quadratic
Fluctuation}

As was shown by Kaneyoshi in \cite{Kan83} the averaged
Green's function $\overline{\G}$ can be expressed approximately 
in terms of mean magnetization
$x$ and its mean quadratic fluctuations $\overline{\xi^2}$. 
To obtain equations
describing these quantities let us consider the equation
for a mean moment at the $l$th site that expresses it in
terms of non--averaged Green's function in the energy representation
$\ll l|l\gg_E$.
In a standard way we obtain
\begin{eqnarray} \label{x_l}
x_l={1\over 2}n_l-\int\limits_{-\infty}^\infty {{\mathrm d} E\over
\ex^{\beta E}-1}
\left(
-{1\over\pi}\Im{\ll l|l\gg}_{E+i 0}
\right).
\end{eqnarray}

Averaging this equation over configuration we obtain the equation for mean
magnetization
\begin{eqnarray} \label{x}
x={1\over 2}n-\int\limits_{-\infty}^\infty {\de E\over
\ex^{\beta E}-1}
\left(
-{1\over\pi N}\Im{{\rm Sp}\overline{\G (E+i 0)}}
\right),
\end{eqnarray}
where we took into account that the averaged one--site Green's function does not
depend on the number of the site:
$$
\overline{\ll l|l\gg}={1\over N}\sum_l\overline{\ll l|l\gg}
={1\over N}\sum_\bq\overline{\ll\bq |\bq\gg}=
{1\over N}{\rm Sp}\overline{\G}.
$$
Subtracting the averaged equation (\ref{x}) from the non--averaged one (\ref{x_l}),
multiplying the result
by the fluctuation $\xi_l$ and averaging the product over configurations we
obtain the following equation for quadratic fluctuations
\begin{eqnarray} \label{xi2_l}
\overline{\xi_l^2}={1\over 2}\overline{\Delta n_l\xi_l}
-\int\limits_{-\infty}^\infty {\de E\over \ex^{\beta E}-1}
\left(
-{1\over\pi}\Im\overline{\xi_l{\ll l|l\gg}_{E+i 0}}
\right).
\end{eqnarray}
It is easy to prove that
$$
\overline{\Delta n_l\xi_l}=
\overline{
(n_l-n)(n_l<S^z_l>-x)
}=(1-n)x
$$
and
$$
\overline{\xi_l\ll l|l\gg}={x\over N}{\rm Sp}\overline{\Gr}.
$$
Substituting these two equations in (\ref{xi2_l}) we obtain finally
\begin{eqnarray} \label{xi2}
\overline{\xi^2}=
x\left(
{1\over 2}(1-n)-\int\limits_{-\infty}^\infty {\de E\over
\ex^{\beta E}-1}
\left(
-{1\over\pi N}\Im{\rm Sp}\overline{\Gr (E+i 0)}
\right)\right),
\end{eqnarray}
Let us consider the meaning of mean quadratic fluctuations $\overline{\xi^2}$
more closely. We find that
\begin{equation}
\overline{\xi^2}=\overline{(n_l<S^z_l>)^2}-x^2=
n\overline{\overline{<S^z_l>^2}}-x^2=
n\left(
\overline{\overline{<S^z_l>^2}}-\sigma^2
\right)
+{1-n\over n}x^2=
n\ds+{1-n\over n}x^2,
\end{equation}
where the first term in the final expression corresponds just to the
fluctuations of mean moment of magnetic atoms while 
the second term corresponds
to the fluctuations of structure. Therefore
it is convenient to divide these two
terms. We can easily find that the mean quadratic fluctuations of moments of
magnetic atoms have the form
\begin{eqnarray} \label{ds0}
\ds=
\sigma\left(
{1\over 2}(1-n)-{1-n\over n}x-
\int\limits_{-\infty}^\infty {\de E\over \ex^{\beta E}-1}
\left(
-{1\over\pi N}\Im{{\rm Sp}\overline{\Gr (E+i 0)}}
\right)\right).
\end{eqnarray}
We can use the equation for mean magnetization (\ref{x}) to rewrite this
equation in the following form
\begin{eqnarray} \label{ds}
\ds=
\sigma
\int\limits_{-\infty}^\infty {\de E\over \ex^{\beta E}-1}
\left(
-{1\over\pi N}\Im{\rm Sp}\overline{\Gs (E+i 0)}
\right),
\end{eqnarray}
where we have introduced the new Green's function
\begin{equation}\label{Gs-def}
\Gs={1-n\over n}\G-\Gr
\end{equation}
The averaged value of this Green's function can be derived from
equations (\ref{G-a}) and
(\ref{Gr-a}) as follows
\begin{equation}\label{Gs-a}
\overline{\Gs}=\left(1-\g\sum_{i=1}^\infty\overline{{\Delta\Q_i\Q}}
\right)^{-1}\g\left(-{\ds\over n\sigma^2}+
{1-n\over n}\sum_{i=1}^\infty\overline{{\Delta\Q_i\rhob}}
-\sum_{i=1}^\infty\overline{{\Delta\Q_i(\rhob+\rhob^2)}}
\right),
\end{equation}
or
\begin{equation}\label{Gs_q-a}
[\overline{\Gs}]_\bq =
{
[\g]_\bq\left(-{\ds\over n\sigma^2}+
{1-n\over n}\sum\limits_{i=1}^\infty[\overline{\Delta\Q_i\rhob}]_\bq 
-\sum\limits_{i=1}^\infty[\overline{\Delta\Q_i(\rhob+\rhob^2)}]_\bq 
\right)
\over
1-[\g]_\bq\sum\limits_{i=1}^\infty[\overline{\Delta\Q_i\Q}]_\bq 
}
\end{equation}
for the diagonal elements in the momentum space.
Here we have used the fact that
\begin{equation}\label{r2}
\overline{\rhob^2}={\overline{\xi^2}\over x^2}={\ds\over n\sigma^2}
+{1-n\over n}.
\end{equation}
It is convenient to come to the dimensionless variables
\begin{eqnarray}
\cE={E-h\over xJ(0)}, \ \ \ \tilde{\beta}=\beta J(0), \ \ \
\tilde h={h\over J(0)}, \ \ \
\Gd_\alpha=J(0)\G_\alpha.
\end{eqnarray}
Then the equations for the magnetization $x$ and
the mean quadratic fluctuations of magnetization of magnetic subsystem $\ds$
take the form
\begin{eqnarray} \label{x-dim}
x&=&{n\over 2}
\left(1+2\int\limits_{-\infty}^\infty {\de \cE\over
\ex^{\tilde{\beta}(\cE x+\tilde{h})}-1}
g(\cE)
\right)^{-1},\\ \label{ds-dim}
{\ds\over n\sigma^2}&=&
2\int\limits_{-\infty}^\infty{\de \cE\over
\ex^{\tilde{\beta}(\cE x+\tilde{h})}-1}
g_2(\cE),
\end{eqnarray}
where
\begin{eqnarray}
g_\alpha(\cE)=
-{1\over 2\pi N}\Im{{\rm Sp}\overline{\Gd_\alpha (\cE+i 0)}}.
\end{eqnarray}

We can assume that the spectral density $g_(\cE)$ does not depend on the mean magnetization $x$,
similarly as in the case of the non--dilute system.
Then substituting zero external field into equation (\ref{x-dim}) 
for magnetization $x$ and looking for the limit of the 
zero magnetization we obtain in a standard way the following equation for critical temperature
\begin{eqnarray}\label{T_c}
{T_{\rm c}\over J(0)}={n\left/ 4\int\limits_{-\infty}^\infty {{\de\cE\over
\cE}g(\cE)}\right.}.
\end{eqnarray}

To solve this set of equations self--consistently we need to express the averaged
Green's functions $\overline{\Gd}$ and
$\overline{\Gsd}$ in terms of
the magnetization $x$ and quadratic fluctuations of the magnetic
subsystem $\ds\over\sigma^2$ only. In the following section we consider
the limit of low concentration of nonmagnetic impurities satisfying this 
condition. The analysis of different approximations allowing to reach
this aim for any concentration of nonmagnetic impurities will be the subject of 
further works.

\setcounter{equation}{0}
\section {Low Concentration of Nonmagnetic Impurities}
Let us estimate the value
\begin{equation}
\overline{
\rho_{\bq,\bk_1}
\rho_{\bk_1,\bk_2}\ldots
\rho_{\bk_{m-1},\bq}
}=\left.{1\over N^m x^m}\sum_{l_1,\ldots,\l_m}
\exp\left({-i\sum_{j=1}^m {\bf l_j}(\bk_j-\bk_{j-1})}\right)
\overline{\xi_{l_1}\ldots\xi_{l_m}}\right|_{\bk_0=\bk_m=\bq}
\end{equation}
we need to calculate to obtain the averaged Green's functions.
For the low concentration of non--magnetic atoms we can neglect the
correlations of fluctuations on different sites supposing
\begin{equation}\label{uncor}
\overline{\xi_{l_1}^{m_1}\ldots\xi_{l_j}^{m_j}}=
\overline{\xi_{l_1}^{m_1}}\ldots\overline{\xi_{l_j}^{m_j}}.
\end{equation}
The only values we need to calculate within this approximation are
$\overline{\xi_l^m}$. It is easy to show that
\begin{equation}\label{xi-m-a}
\overline{\xi_l^m}=(1-n)(-x)^m+n\overline{\overline{
(\Delta\sigma+(1-n)\sigma)^m
}}=(1-n)(-x)^m+n\overline{\overline{
(\Delta\sigma)^m
}}+o(1-n).
\end{equation}

Let us remind that for the Gauss distribution of the random value $\zeta$
\begin{equation}
\overline{(\Delta \zeta)^{2m}}=(2m-1)!!\overline{(\Delta \zeta)^{2}}^m, \ \ \ \ \
\overline{(\Delta \zeta)^{2m+1}}=0.
\end{equation}
We don't now the distribution law for magnetization fluctuations
but we can suppose that the other than quadratic fluctuations are 
small enough to be neglected within the linear as to the concentration of
non--magnetic atoms approximation
\begin{equation}
\overline{
        \overline{(\Delta \sigma)^m}}=
o\left(\overline{\overline{(\Delta \sigma)^2}}\right), \ \ \ m>2.
\end{equation}
Hereafter we shall neglect higher than quadratic moments
of magnetization of the magnetic subsystem.
Within the made approximation we obtain
\begin{eqnarray}\label{arhorho}
\overline{
\rho_{\bq,\bk_1}
\rho_{\bk_1,\bk_2}\ldots
\rho_{\bk_{m-1},\bq}
}
\approx
{1\over N^{m-1}}{\overline{\left({\xi\over x}\right)^m}}+
\\ \nonumber
{1\over N^{m-2}}\sum_{j=2}^{m-2}
{\overline{\left({\xi\over x}\right)^j}}\
{\overline{\left({\xi\over x}\right)^{m-j}}}
\sum\limits_{\{\lambda\}_{m-1}^{j-1}}
\left.
\deltaF_{\rm K}\left(\bq -\bk_{m-1}+
\sum\limits_{f=1}^{j-1}(\bk_{\lambda_f}-\bk_{\lambda_f-1})\right)
\right|_{\bk_0=\bq},
\end{eqnarray}
where
$\{\lambda\}_m^j\equiv{\{\lambda_1,\ldots,\lambda_j\}\subset\{1,\ldots,m\}}$
and
$\deltaF_{\rm K}(x)\equiv\deltaF_{x,0}$ is Kronecker's symbol. The first term in 
the right hand side of expression (\ref{arhorho}) corresponds to the one--site 
approximation and the second one corresponds to the two--site approximation
where we have neglected possible correlations of the fluctuations on 
different sites.

Let us calculate the Green's functions
$\overline{\G}$ and
$\overline{\Gs}$ taking into account only the terms
linear as to the
concentration of nonmagnetic atoms. Within the made approximations the
series in expressions (\ref{G_q-a}) and (\ref{Gs_q-a}) for the
averaged Green's functions $[\overline{\G}]_\bq $ and
$[\overline{\Gs}]_\bq $ can be summed up.
Indeed
\begin{eqnarray}
\sum\limits_{i=1}^\infty[\overline{\Delta\Q_i\rhob}]_\bq =
\sum\limits_{i=1}^\infty[\overline{(\Q\g)^i\rhob}]_\bq +o(1-n)&\approx&\\
\sum\limits_{i=1}^\infty{N^i\ \sum\limits_{\bk_1,\ldots,\bk_i}}
B_{\bq,\bk_1}
B_{\bk_1,\bk_2}\ldots
B_{\bk_{i-1},\bk_i}
\ \overline{
\rho_{\bq,\bk_1}
\rho_{\bk_1,\bk_2}\ldots
\rho_{\bk_i,\bq}}&\approx&
\sum\limits_\bk\sum\limits_{i=1}^\infty
\overline{{\left(\xi\over x\right)}^{i+1}}
[\bB^i]_{\bf q,k},\nonumber
\end{eqnarray}
where
$$
B_{\bf q,k}={1\over N}{A_{\bf q,k}\over\cE-\cE_0(\bk )}, \ \ \ \ \
A_{\bf q,k}=\tilde{J}({\bf q-k})-\tilde{J}(\bk ), \ \ \ \ \
\cE_0(\bq )=1-\tilde{J}(\bq ), \ \ \ \ \
\tilde{J}(\bq )={J(\bq )\over J(0)}.
$$
Taking into account the expression for the averaged
fluctuations $\overline{\xi^m}$ (\ref{xi-m-a}) we get
\begin{eqnarray}\label{ser1}
\sum\limits_{i=1}^\infty[\overline{\Delta\Q_i\rhob}]_\bq &\approx&
\sum\limits_\bk\sum\limits_{i=1}^\infty
\left((1-n)(-1)^{i+1}+n{\overline{\overline{
(\Delta\sigma)^{i+1}
}}\over x^{i+1}}
\right)[\bB^i]_{\bf q,k}\approx 
\\ &&\nonumber
(1-n)\sum\limits_\bk [\bB({\bf 1+B})^{-1}]_{\bf q,k}+
{\ds\over \sigma^2}\sum\limits_\bk B_{\bf q,k}.
\end{eqnarray}
In a similar way we can estimate other series as follows
\begin{eqnarray}\label{ser2}
\sum\limits_{i=1}^\infty[\overline{\Delta\Q_i\Q}]_\bq\approx
{1\over 2}\left(
(1-n)[\bB({\bf 1+B})^{-1}\bA]_{\bf q,q}+
{\ds\over \sigma^2}[{\bf BA}]_{\bf q,q}
\right)
\end{eqnarray}
and
\begin{eqnarray}\label{ser3}
\sum\limits_{i=1}^\infty[\overline{\Delta\Q_i\rhob^2}]_\bq\approx
-(1-n)\sum\limits_\bk [\bB({\bf 1+B})^{-1}]_{\bf q,k}.
\end{eqnarray}
Thus we obtain the approximat expressions for the averaged Green's
functions $\overline{\Gd}$ and
$\overline{\Gsd}$ in terms of the mean magnetization $x$ and structural
magnetic fluctuations of magnetic subsystem $\ds\over \sigma^2$ only.
\begin{equation}\label{G_q-a-low}
[\overline{\Gd}]_\bq =2
{1+
{\ds\over \sigma^2}\sum\limits_\bk B_{\bf q,k}+
(1-n)\sum\limits_\bk [\bB({\bf 1+B})^{-1}]_{\bf q,k}
\over
\cE-\cE_0(\bq )-
{\ds\over \sigma^2}[{\bf BA}]_{\bf q,q}-
(1-n)[\bB({\bf 1+B})^{-1}\bA]_{\bf q,q}
}
\end{equation}
and
\begin{equation}\label{Gs_q-a-low}
[\overline{\Gsd}]_\bq =2
{-
{\ds\over \sigma^2}\left(1+\sum\limits_\bk B_{\bf q,k}\right)
\over
\cE-\cE_0(\bq )-
{\ds\over \sigma^2}[{\bf BA}]_{\bf q,q}-
(1-n)[\bB({\bf 1+B})^{-1}\bA]_{\bf q,q}
}.
\end{equation}
Substituting expression (\ref{Gs_q-a-low}) in equation (\ref{ds-dim})
we can find that it gives us zero solution for the
fluctuations $\ds\over\sigma^2$. It means that this value is of a higher order
as to concentration of non--magnetic impurities and it should be
neglected within the linear approximation. Thus we can rewrite the Green's
function $\overline{\Gd}$ in the form
\begin{equation}\label{G_q-a-lin}
[\overline{\Gd}]_\bq =2
{1+
(1-n)C(\cE,\bq )
\over
\cE-\cE_0(\bq )-
(1-n)
\Sigma(\cE,\bq )
},
\end{equation}
where
\begin{eqnarray}
C(\cE,\bq )=\sum\limits_\bk [\bB({\bf 1+B})^{-1}]_{\bf q,k},
\label{C}\\
\Sigma(\cE,\bq )=[\bB({\bf 1+B})^{-1}\bA]_{\bf q,q}.
\label{Sigma}
\end{eqnarray}
This expression contains neither magnetization
nor its fluctuations. The problem of calculating the inverse matrix
$({\bf 1+B})^{-1}$ is well known in theory of crystals with impurities
and similar matrices were calculated in \cite{Kan69.42} and \cite{Iz}.
We present a simple way of calculating the functions $C$ and $\Sigma$
for the $d$--dimensional simple cubic lattice with the nearest neighbours
interaction in appendix A.

From the poles of the averaged function $\overline{\Gd}$ we can obtain
the equation for the spectrum of spin excitations
\begin{equation}\label{Spectrum}
\cE-\cE_0(\bq )-
(1-n)
\Sigma(\cE,\bq )=0
\end{equation}
We present here only the long--wave solution of this equation for the simple
cubic lattice with the nearest neighbours interaction
\begin{eqnarray}\label{SpectrumL}
E\approx D q^2=x J a^2\left(1-
2(1-n)
{I_{s^2}(d)\over 1-I_{s^2}(d)}
\right)q^2=
\\ \nonumber
\sigma J a^2 n \left(1-
2(1-n)
{I_{s^2}(d)\over 1-I_{s^2}(d)}
\right)q^2=
\sigma J a^2 \left(1-
(1-n)
{1+I_{s^2}(d)\over 1-I_{s^2}(d)}
+o[(1-n)^2]\right)q^2,
\end{eqnarray}
where $J$ is the exchange integral for the nearest sites, $a$ is the lattice
constant, $d$ is the dimensionality of lattice. The integral
\begin{equation}\label{Id}
I_{s^2}(d)={1\over(2\pi)^d}\int\limits_{-\pi}^{\pi}\de x_1\ldots\int\limits_{-\pi}^{\pi}\de x_d
{\sin^2(x_1)\over\sum\limits_{\mu=1}^d (1-\cos(x_\mu))}
\end{equation}
is smaller than the one for all the dimensions except $d=1$. In the one--dimensional
case the denominator of the second term in the expression for spectrum
(\ref{SpectrumL}) diverges
that corresponds to the fact that spin--waves can not
propagate in the one--dimensional ferromagnet with the nearest
neighbours interaction at any dilution. 
For the dimensions $d\ge 2$ the spin--wave stiffness
constant $D$ in expression (\ref{SpectrumL}) tends to zero while the 
concentration of the magnetic sites $n$ decreases to the percolation
limit $n_{\rm c}$ which within the linear as to the concentration of nonmagnetic 
impurities approximation takes the form
\begin{equation}\label{nc}
n_{\rm c}=1-{1-I_{s^2}(d)\over 1+I_{s^2}(d)}={2 I_{s^2}(d)\over 1+I_{s^2}(d)}
\end{equation}

Note that we can reduce integral (\ref{Id}) to the one--dimensional form  
\begin{equation}
I_{s^2}(d)=\int\limits_{0}^\infty
{\displaystyle
\frac {{\rm e}^{-d\,t}\,I_1(t)\,I_0^{d-1}(t)}{t}
}\,
\de t,
\end{equation}
where we have used the transformation
\begin{equation}
{1\over\sum\limits_{\mu=1}^d (1-\cos(x_\mu))}=
\int\limits_{0}^\infty \exp{\left(-t\sum\limits_{\mu=1}^d (1-\cos(x_\mu))
\right)}\de t
\end{equation}
and the following properties of the modified Bessel functions $I_\nu(t)$
\begin{equation}
{1\over 2\pi}
\int\limits_{-\pi}^{\pi}\ex^{t\cos x}\cos(nx)\de x= I_n(t), \ \ \ 
{1\over 2\pi}
\int\limits_{-\pi}^{\pi}{\rm e}^{t\cos x} \sin^2 x \ \de x=
 {I_0(t)-I_2(t)\over 2}={I_1(t)\over t}.
\end{equation}
As the expression for the percolation threshold (\ref{nc}) is obtained within the linear 
over $1-n$ approximation it shall give reliable results for low dimensions when 
$1-n_{\rm c}$ is small but we can not expect any reliability for large $d$ when 
$(1-n_{\rm c})\to 0$. Indeed, for $d\to\infty$ the percolation threshold 
$n_{\rm c}\to {1\over d}$ that does 
not coincide with the Bethe expression $n_{\rm c}^{\rm B}= {1\over 2 d-1}$ which is an 
asymptotic of the percolation threshold at large $d$. Nevertheless as we can 
see from the Table~I the results for the percolation threshold obtained from 
(\ref{nc}) are better for $d=2$ and $d=3$ than the ones obtained 
within the Bethe--lattice approach.
One can find some results of the latter
approach for the theory of dilute ferromagnets in
\cite{Salz76.18, Salz76.14, Du}.

Table I.
{\small
Percolation thresholds of the $sc$ lattice from this work $n_{\rm c}$
in comparison with the "exact estimates"
$n_{\rm c}^{\rm e}$ taken from \cite{per} and Bethe expression $n_{\rm c}^{\rm B}=1/(q-1)$. 
For the $sc$
lattice the number of the nearest neighbours $q=2d$.
}
\par\nopagebreak
\bigskip
\nopagebreak
\begin{center}
\begin{tabular}
{|c@{~~\vline}c@{~~\vline}c@{~~\vline}c@{~~}|}\hline
&&&\\
{~Dimension~}&{~~~$n_{\rm c}^{\rm e}$~~~}&{~~~$n_{\rm c}^{\rm B}$~~~}&{~~~$n_{\rm c}$~~~}\\
&&&\\
\hline
{~$d=1$~}&{~~1~~}&{~~1~~}&{~~1~~}\\
{~$d=2$~}&{~0.5928~}&{~~0.(3)~~}&{~0.53306~}\\
{~$d=3$~}&{~0.3116~}&{~~0.2~~}&{~0.34689~}\\
{~$d=4$~}&{~0.197~}&{~0.(142857)~}&{~0.25585~}\\
{~$d=5$~}&{~0.141~}&{~~0.(1)~~}&{~0.20286~}\\
{~$d=6$~}&{~0.107~}&{~~0.(09)~~}&{~0.16824~}\\
{~$d=7$~}&{~0.089~}&{~~0.(076923)~~}&{~0.14380~}\\
{~$d=8$~}&{~~}&{~~0.0(6)~~}&{~0.12561~}\\
{~$d=9$~}&{~~}&{~~0.058823~~}&{~0.11153~}\\
{~$d=10$~}&{~~}&{~0.052632~}&{~0.10030~}\\
&&&\\
{~$d=100$~}&{~~}&{~~0.0050251~~}&{~0.01000~}\\
\hline
\end{tabular}
\end{center}

Let us write an explicit expression for the spectral density $g(\cE)$ which we need 
to calculate magnetization (\ref{x-dim}) and critical temperature (\ref{T_c})
\begin{eqnarray}
g(\cE)&=&{1 \over N}\sum_\bq g(\cE,\bq ),\\\nonumber
g(\cE,\bq )\equiv-{1\over 2\pi}\Im[\overline{\Gd}]_\bq 
(\cE+i0)&=&
{1\over \pi}{{(1-n)\Sigma''(\cE,\bq )+F(\cE,\bq )}
\over
\left(\cE-\cE_0(\bq )-(1-n)\Sigma'(\cE,\bq )
\right)^2+\left((1-n)\Sigma''(\cE,\bq )\right)^2},
\end{eqnarray}
where
\begin{eqnarray}
F(\cE,\bq )=
(1-n) C''(\cE,\bq )
\left(
        \cE-\cE_0(\bq )-(1-n)\Sigma'(\cE,\bq )
\right)
+(1-n)^2
C'(\cE,\bq )
\Sigma''(\cE,\bq ).
\end{eqnarray}
The functions $C'$, ${\Sigma}'$ and $C''$, ${\Sigma}''$ are respectively 
the real and
the imaginary parts of the corresponding functions (\ref{C}) and 
(\ref{Sigma}). They are defined as follows
\begin{equation}\label{R-I}
C(\cE+i0,\bq )=C'(\cE,\bq )-iC''(\cE,\bq ),\ \ \
\Sigma(\cE+i0,\bq )=
\Sigma'(\cE,\bq )-i\Sigma''(\cE,\bq ).
\end{equation}
We can see that the expression for the obtained spectral density $g(\cE)$ does 
not depend on the magnetization and its fluctuations. It confirms the assumption 
we have made in the second section to obtain 
expression (\ref{T_c}) for the critical temperature. 

Let us estimate now the structural fluctuations of magnetization of a magnetic
subsystem at low concentration of non--magnetic impurities. To do it we assume
that these fluctuations are proportional to the square of concentration of
non--magnetic atoms and we shall keep in the numerator of the expression
(\ref{Gs_q-a}) for the Green's function $\Gs$ all the terms up to the second
order as to the concentration $(1-n)$ of non--magnetic sites.

Thus we need to sum up the series
\begin{eqnarray}
\sum\limits_{i=1}^\infty[\overline{\Delta\Q_i(\rhob+\rhob^2)}]_\bq =
\sum\limits_{i=1}^\infty[\overline{\Q_i(\rhob+\rhob^2)}]_\bq -
\sum\limits_{i=2}^\infty[\overline{\Q_i}][\overline{\rhob^2}]_\bq =
\\ \nonumber
\left(
1-\sum\limits_{j=2}^\infty[\overline{(\Q\g)^j}]_\bq 
\right)
\sum\limits_{i=1}^\infty[\overline{(\Q\g)^i(\rhob+\rhob^2)}]_\bq -
\sum\limits_{i=2}^\infty[\overline{\Q_i}][\overline{\rhob^2}]_\bq +
o[(1-n)^2]
\end{eqnarray}
As we can see from expressions (\ref{ser1}) and (\ref{ser3}) the series
$\sum\limits_{i=1}^\infty[\overline{(\Q\g)^i(\rhob+\rhob^2)}]_\bq $
is of
the same order as the fluctuations $\ds\over\sigma^2$ and therefore it is
within the made assumptions of the quadratic order with respect to the
concentration of non--magnetic impurities. Thus we have
\begin{eqnarray}\label{rr2}
&&
\sum\limits_{i=1}^\infty[\overline{\Delta\Q_i(\rhob+\rhob^2)}]_\bq\approx
\sum\limits_{i=1}^\infty[\overline{(\Q\g)^i(\rhob+\rhob^2)}]_\bq-
\sum\limits_{i=2}^\infty[\overline{\Q_i}][\overline{\rhob^2}]_\bq=
\\&& \nonumber
N \sum\limits_\bk 
B_{\bq,\bk}
\left[
\ \overline{
\rho_{\bq,\bk}
\rho_{\bk,\bq}
}
+
\sum\limits_{\bk'}
\overline{
\rho_{\bq,\bk}
\rho_{\bk,\bk'}
\rho_{\bk',\bq}
}
\right]+
\sum\limits_{i=2}^\infty{N^i\ \sum\limits_{\bk_1,\ldots,\bk_i}}
B_{\bq,\bk_1}
B_{\bk_1,\bk_2}\ldots
B_{\bk_{i-1},\bk_i}\times
\\&& \nonumber
\left[
\ \overline{
\rho_{\bq,\bk_1}
\rho_{\bk_1,\bk_2}\ldots
\rho_{\bk_i,\bq}}
+
\sum_{\bk_{i+1}}
\left(
\overline{
\rho_{\bq,\bk_1}
\rho_{\bk_1,\bk_2}\ldots
\rho_{\bk_{i+1},\bq}
}-\deltaF_{\bk_i,\bq}\
\overline{
\rho_{\bq,\bk_1}
\rho_{\bk_1,\bk_2}\ldots
\rho_{\bk_{i-1},\bq}
}\
\overline{
\rho_{\bq,\bk_{i+1}}
\rho_{\bk_{i+1},\bq}
}
\right)
\right]\approx
\end{eqnarray}
\begin{eqnarray} \label{ser5}\nonumber
&&\left[
\overline{\left({\xi\over x}\right)^2}+
\overline{\left({\xi\over x}\right)^3}
\right]
\sum\limits_\bk 
B_{\bq,\bk}+
\sum\limits_{i=2}^\infty\sum\limits_{\bk_1,\ldots,\bk_i}
B_{\bq,\bk_1}
B_{\bk_1,\bk_2}\ldots
B_{\bk_{i-1},\bk_i}\times\\\nonumber
&&\left\{
\overline
{
\left(
{\xi\over x}
\right)^{i+1}
}+
N\sum_{j=2}^{i-1}
\overline{\left({\xi\over x}\right)^j}\
\overline{\left({\xi\over x}\right)^{i+1-j}}
\sum\limits_{\{\lambda\}^{j-1}_i}
\deltaF\left(\bq -\bk_i+
\sum\limits_{f=1}^{j-1}(\bk_{\lambda_f}-\bk_{\lambda_f-1})\right)+
\right.\\\nonumber
&&\left.
{1\over N}\sum_{\bk_{i+1}}\left[
\overline{\left({\xi\over x}\right)^{i+2}}+
N\sum_{j=2}^i
\overline{\left({\xi\over x}\right)^j}\
\overline{\left({\xi\over x}\right)^{i+2-j}}
\sum\limits_{\{\lambda\}^{j-1}_{i+1}}
\deltaF\left(\bq -\bk_{i+1}+
\sum\limits_{f=1}^{j-1}(\bk_{\lambda_f}-\bk_{\lambda_f-1})\right)
\right]-\right.\\\nonumber
&&\left.
N
\overline{\left({\xi\over x}\right)^i}\
\overline{\left({\xi\over x}\right)^2}
\deltaF_{\bk_i,\bq}
\right\}=
\end{eqnarray}
\begin{eqnarray}
\nonumber
&&\sum\limits_{i=1}^\infty \sum\limits_\bk 
[\bB^i]_{\bq,\bk}
\left[
\overline{\left({\xi\over x}\right)^{i+1}}+
\overline{\left({\xi\over x}\right)^{i+2}}
\right]+
\sum\limits_{i=2}^\infty\sum\limits_{\bk_1,\ldots,\bk_i}
B_{\bq,\bk_1}
B_{\bk_1,\bk_2}\ldots
B_{\bk_{i-1},\bk_i}\times\\\nonumber
&&\left\{
N\sum_{j=2}^{i-1}
\overline{\left({\xi\over x}\right)^j}\
\overline{\left({\xi\over x}\right)^{i+1-j}}
\sum\limits_{\{\lambda\}^{j-1}_i}
\deltaF\left(\bq -\bk_i+
\sum\limits_{f=1}^{j-1}(\bk_{\lambda_f}-\bk_{\lambda_f-1})\right)+
\right.\\\nonumber
&&\left.
\sum_{\bk_{i+1}}
\sum_{j=2}^i
\overline{\left({\xi\over x}\right)^j}\
\overline{\left({\xi\over x}\right)^{i+2-j}}
\sum\limits_{\{\lambda\}^{j-1}_{i+1}}
\deltaF\left(\bq -\bk_{i+1}+
\sum\limits_{f=1}^{j-1}(\bk_{\lambda_f}-\bk_{\lambda_f-1})\right)-
%\right.\\\nonumber &&\left.
N
\overline{\left({\xi\over x}\right)^i}\
\overline{\left({\xi\over x}\right)^2}
\deltaF_{\bk_i,\bq}
\right\}.
\end{eqnarray}
It is easy to show that
\begin{eqnarray}\label{red}
&&\sum_{\bk_{i+1}}
\sum\limits_{\{\lambda\}^{j-1}_{i+1}}
\deltaF\left(\bk_0-\bk_{i+1}+
\sum\limits_{f=1}^{j-1}(\bk_{\lambda_f}-\bk_{\lambda_f-1})\right)=
\\ \nonumber
&&\sum_{\bk_{i+1}}
\sum\limits_{\{\lambda\}^{j-2}_i}
\deltaF\left(\bk_0-\bk_i+
\sum\limits_{f=1}^{j-2}(\bk_{\lambda_f}-\bk_{\lambda_f-1})\right)+
\sum_{\bk_{i+1}}
\sum\limits_{\{\lambda\}^{j-1}_i}
\deltaF\left(\bk_0-\bk_{i+1}+
\sum\limits_{f=1}^{j-1}(\bk_{\lambda_f}-\bk_{\lambda_f-1})\right)=
\\ \nonumber
&&N\sum\limits_{\{\lambda\}^{j-2}_i}
\deltaF\left(\bk_0-\bk_i+
\sum\limits_{f=1}^{j-2}(\bk_{\lambda_f}-\bk_{\lambda_f-1})\right)+
C^{j-1}_i,
\end{eqnarray}
where $C^m_n={n!\over m!(n-m)!}$ is the binomial coefficient
indicating the number of different $m$--element subsets of
the set $\{1,\ldots,n\}$. Substituting (\ref{red}) in the expression (\ref{rr2}) we
obtain
\begin{eqnarray}
&&\sum\limits_{i=1}^\infty
[\overline{\Delta\Q_i(\rhob+\rhob^2)}]_\bq\approx
\\ \nonumber
&&\sum\limits_{i=1}^\infty\sum\limits_\bk 
[\bB^i]_{\bq,\bk}
\left(
\overline{\left({\xi\over x}\right)^{i+1}}+
\overline{\left({\xi\over x}\right)^{i+2}}
\right)+
%\\ \nonumber &&
\sum\limits_{i=2}^\infty\sum\limits_\bk 
[\bB^i]_{\bq,\bk}
\left(
\sum_{j=1}^{i-1}
\overline{\left({\xi\over x}\right)^{j+1}}\
\overline{\left({\xi\over x}\right)^{i+1-j}}
{i!\over j!(i-j)!}\right)
+
\\\nonumber
&&N
\sum\limits_{i=2}^\infty
\sum\limits_{\bk_1,\ldots,\bk_i}
B_{\bq,\bk_1}
\ldots
B_{\bk_{i-1},\bk_i}
\sum_{j=1}^{i-2}
\overline{\left({\xi\over x}\right)^{i-j}}
\left(
\overline{\left({\xi\over x}\right)^{j+1}}+
\overline{\left({\xi\over x}\right)^{j+2}}
\right)
%\times\\ \nonumber &&
\sum\limits_{\{\lambda\}^j_i}
\left.
\deltaF\left(\bq -\bk_i+
\sum\limits_{f=1}^j(\bk_{\lambda_f}-\bk_{\lambda_f-1})\right)
\right|_{\bk_0=\bq}
\end{eqnarray}
Now we substitute in this expression the value of
$\overline{\left({\xi\over x}\right)^m}$ from equation (\ref{xi-m-a}) 
and finally we obtain
\begin{eqnarray}
\sum\limits_{i=1}^\infty
[\overline{\Delta\Q_i(\rhob+\rhob^2)}]_\bq&\approx&
\left(
{\ds\over\sigma^2}
+(1-n)^2
\right)
\sum_\bk B_{\bq,\bk}+
(1-n)^2 
\sum\limits_{i=2}^\infty\sum\limits_\bk 
[\bB^i]_{\bq,\bk}
(-1)^i(2^i-2)=
\\\nonumber
&&\left(
{\ds\over\sigma^2}
+(1-n)^2
\right)
\sum_\bk B_{\bq,\bk}+
(1-n)^2 
\sum_\bk [4\bB^2(1+2\bB)^{-1}-
2\bB^2(1+\bB)^{-1}]_{\bf q,k},
\end{eqnarray}
where we have used the identity
\begin{equation}
\sum_{j=1}^{i-1}
{i!\over j!(i-j)!}
\equiv 2^i-2.
\end{equation}
Finally we get the following expression for the Green's function $\Gsd$
\begin{equation}\label{Gs_q-a-low2}
\overline{[\Gsd]}_\bq =2
{
{\ds\over \sigma^2}\left(1+\sum\limits_\bk B_{\bf q,k}\right)
+(1-n)^2\sum\limits_\bk [\bB+(1+\bB)^{-1}-(1+2\bB)^{-1}]_{\bf q,k}
\over
\cE-\cE_0(\bq )
},
\end{equation}
where we have neglected the contribution of disorder to the denominator 
because of both the terms in the numerator are of the order of the square of
the concentration of non--magnetic impurities.
It is convenient to distincguish the two parts of this Green's function as follows
\begin{equation}\label{Gs-ab}
\overline{\Gs}=(1-n)^2\Gsa-{\ds\over \sigma^2}\Gsb,
\end{equation}
where
\begin{eqnarray}\label{Gsab_q}
[\Gsad]_\bq =2
{
C_a(\cE,\bq )
\over
\cE-\cE_0(\bq )
},
\ \ \
[\Gsbd]_\bq =2
{
1+C_b(\cE,\bq )
\over
\cE-\cE_0(\bq )
},\\\label{Cab}
C_a(\cE,\bq )=
\sum\limits_\bk [\bB+(1+\bB)^{-1}-(1+2\bB)^{-1}]_{\bf q,k},
\ \ \
C_b(\cE,\bq )=
\sum\limits_\bk B_{\bf q,k}.
\end{eqnarray}

Then we can rewrite the equations for the mean quadratic fluctuations of 
the magnetization of the magnetic subsystem $\ds$ (\ref{ds-dim}) in the following
form
\begin{eqnarray}\label{ds-dim2}
{\ds\over\sigma^2}=
2(1-n)^2\int\limits_{-\infty}^\infty{\de \cE\over
\ex^{\tilde{\beta}(\cE x+\tilde{h})}-1}
g_{2a}(\cE)\left/
\left(
1+2\int\limits_{-\infty}^\infty{\de \cE\over
\ex^{\tilde{\beta}(\cE x+\tilde{h})}-1}
g_{2b}(\cE)
\right)
\right.,
\end{eqnarray}
where
\begin{eqnarray}\label{ga}
g_{2a}(\cE)\equiv-{1\over 2\pi N}{\rm Sp}\Im\overline{\Gsad},&&
g_{2b}(\cE)\equiv-{1\over 2\pi N}{\rm Sp}\Im\overline{\Gsbd}.
\end{eqnarray}

Let us consider the low--temperature behaviour of structure fluctuations of 
the magnetization of the magnetic subsystem $\ds\over\sigma^2$. Both the integrals in 
equation(\ref{ds-dim2}) tend to zero at the temperature tending to zero. 
Therefore we need only the low--energy behaviour of the spectral density 
$g_{2a}$ to obtain the low--temperature asymptotic of the fluctuations 
$\ds\over\sigma^2$. In the case of the simple cubic
lattice the simple calculations shown in the appendix B yield
\begin{equation}g_{2a}(\cE)\mathop{\to}\limits_{T\to 0}{2\over N}
\sum_\bq\deltaF(\cE-\cE_0(\bq ))
\end{equation}
giving in the three--dimensional case
\begin{equation}
{\ds\over\sigma^2}=4(1-n)^2 Z_{3/2}(\beta h)
\left(T\over 2\pi n J\right)^{3/2},
\end{equation}
where
$$
Z_\nu(x)=\sum_{n=1}^\infty{n^{-\nu}\ex^{-nx}}
$$
and $J$ is the exchange integral for the nearest neighbours.

\setcounter{equation}{0}
\section {Conclusions}
We have shown that the configurationally
averaged Green's function method is a useful tool for deriving
self--consistent equations describing magnetization and its mean
quadratic fluctuations caused by disorder.
In particular, at low concentration of nonmagnetic
impurities we obtain explicit expressions for the spectral densities
$g(\cE)$ and $g_2(\cE)$ giving the value of magnetization $x$
and its quadratic structural fluctuations $\ds$ with respect to 
the linear concentration of impurities.

We have revealed that the relative quadratic fluctuations of
magnetization $\ds\over \sigma^2$ could be neglected within the linear 
as to the concentration of nonmagnetic impurities $1-n$. 
Thay are found to be quadratic over the $1-n$ and indicate the $T^{3/2}$
behaviour at low temperatures.
 
The same approach can be applied to describe
amorphous and liquid spin systems. It will be the subject of
further works.

\begin{appendix}
\renewcommand{\theequation}{\thesection.\arabic{equation}}
\setcounter{equation}{0}
\section{Calculation of the Functions $C$ and $\Sigma$
for the Simple Cubic Lattice}
It is easy to show that the matrix element
$$
B_{\bf q,k}={1\over N}{A_{\bf q,k}\over\cE-\cE_0(\bk )}
$$
could be rewritten as
$$
B_{\bf q,k}={1\over N}
{
\sum\limits_{\bf R}\tilde{J}({\bf R})
\left[
(\cos({\bf qR})-1
)\cos({\bf kR})+
\sin({\bf qR})\sin({\bf kR})
\right]
\over
\left[
\cE-
\left(1-\sum\limits_{\bf R}
\tilde{J}({\bf R})
\cos({\bf kR})
\right)
\right]
}.
$$
For the nearest--neighbour interaction it takes the following form
\begin{equation}\label{Bnear}
B_{\bf q,k}={1\over N}
{
{1\over z}\sum\limits_{\bf l}
\left[
(\cos({\bf ql})-1
)\cos({\bf kl})+
\sin({\bf ql})\sin({\bf kl})
\right]
\over
\left[
\cE-
\left(1-{1\over z}\sum\limits_{\bf l}
\cos({\bf kl})
\right)
\right]
},
\end{equation}
where ${\bf l}$ goes over the $z$ nearest neighbours of some site.
The fact that the matrix element $B_{\bf q,k}$ is invariant under the
transformation ${\bf l}\to -{\bf l}$ allows us to reduce all the sums in
expression (\ref{Bnear}) as follows
$$
\sum\limits_{\bf l}=
2\sum\limits_{\bf l}',
$$
where a prime denotes that the sum is taken over the $z/2$ nearest neighbours
of some site within one half--space.
Expression (\ref{Bnear}) can be easily factorized as follows
$$
B_{\bf q,k}=\mathop{\sum\limits_{\bf l}'}\limits_{\mu=1, 2}
\alpha_{{\bf l},\mu}(\bq )
\beta_{{\bf l},\mu}(\cE,\bk ),
$$
where
\begin{eqnarray}
\alpha_{{\bf l},1}(\bk )=
{2\over z}
\left(
\cos({\bf kl})-1
\right),&&
\alpha_{{\bf l},2}(\bk )=
{2\over z}
\sin({\bf kl}),\\
\nonumber
\beta_{{\bf l},1}(\cE,\bk )=
{
\cos({\bf kl})
\over
N
\left[
\cE-
\left(1-{2\over z}\sum\limits_{\bf l}'
\cos({\bf kl})
\right)
\right]
},
&&
\beta_{{\bf l},2}(\cE,\bk )=
{
\sin({\bf kl})
\over
N
\left[
\cE-
\left(1-{2\over z}\sum\limits_{\bf l}'
\cos({\bf kl})
\right)
\right]
}.
\end{eqnarray}
Now we can rewrite $B_{\bf q,k}$ as a matrix product
$$
B_{\bf q,k}=\alphab^T(\bq )\betab(\cE,\bk ),
$$
where
\begin{eqnarray}
\alphab=
\left(
\matrix{\alphab_1\cr\alphab_2}
\right)=
\left(
\matrix{
\alpha_{{\bf l}_1,1}\cr
\vdots\cr
\alpha_{{\bf l}_{z/2},1}\cr
\alpha_{{\bf l}_1,2}\cr
\vdots\cr
\alpha_{{\bf l}_{z/2},2}
}
\right),
&&
\betab=
\left(
\matrix{\betab_1\cr\betab_2}
\right)=
\left(
\matrix{
\beta_{{\bf l}_1,1}\cr
\vdots\cr
\beta_{{\bf l}_{z/2},1}\cr
\beta_{{\bf l}_1,2}\cr
\vdots\cr
\beta_{{\bf l}_{z/2},2}
}
\right)
\end{eqnarray}

After the made transformation the inverse matrix $({\bf 1+B})^{-1}$ can be
easily calculated. Indeed
\begin{eqnarray}
[({\bf 1+B})^{-1}]_{\bf q,k}&=&
\deltaF_{\bf q,k}+ \sum_{i=1}^\infty[(-\bB)^i]_{\bf q,k}=\\
\nonumber
&&\deltaF_{\bf q,k}-\alphab^T(\bq )
\left(1+\sum_{i=1}^\infty\left(-\sum_{\bf k'}\betab(\cE,{\bf k'})
\alphab^T({\bf k'})\right)^i\right) \betab(\cE,\bk )=\\
\nonumber
&&\deltaF_{\bf q,k}-\alphab^T(\bq )
(1+\gammab(\cE))^{-1}
\betab(\cE,\bk ),
\end{eqnarray}
where
\begin{equation}
\gammab(\cE)=\sum_\bk\betab(\cE,\bk )
\alphab^T(\bk ).
\end{equation}
Thus the problem of calculating the inverse matrix $({\bf 1+B})^{-1}$ of the order $N$
is reduced to the calculation of inverse matrix $(1+\gammab(\cE))^{-1}$
of the order $z/2$.

Since $\alphab_1$ and $\betab_1$ are even functions of $\bk $ and
$\alphab_2$ and $\betab_2$ are the odd ones the matrix $\gammab$ has the form
\begin{equation}
\gammab(\cE)=
\left(
\matrix
{
\gammab_1(\cE)&0\cr
0&\gammab_2(\cE)
}
\right),
\end{equation}
where
\begin{eqnarray}
\gammab_\nu(\cE)=\sum_\bk\betab_\nu(\cE,\bk )
\alphab_\nu^T(\bk ).
\end{eqnarray}
Now we can easily express the functions $C$ defined in (\ref{C}), (\ref{Cab}) in
terms of the matrices $\alphab$, $\betab$ and $\gammab$
\begin{eqnarray}
C(\cE,\bq )&=&\sum\limits_\bk [\bB({\bf 1+B})^{-1}]_{\bq,\bk}=
%\nonumber\\&&
\alphab^T(\bq )
(1+\gammab(\cE))^{-1}
\betac(\cE),\\
C_a(\cE,\bq )&=&
\sum\limits_\bk [\bB+(1+\bB)^{-1}-(1+2\bB)^{-1}]_{\bf q,k}=
\\\nonumber&&
\alphab^T(\bq )
\left[
1+
2(1+2\gammab(\cE))^{-1}
-(1+\gammab(\cE))^{-1}
\right]
\betac(\cE),\\
C_b(\cE,\bq )&=&
\sum\limits_\bk B_{\bf q,k}=
%\nonumber\\&&
\alphab^T(\bq )
\betac(\cE),
\end{eqnarray}
where
\begin{equation}
\betac(\cE)=\sum\limits_\bk\betab(\cE,\bk ).
\end{equation}
To obtain a similar expression for the function $\Sigma$ we rewrite the
matrix $\bA$ in a similar way as the $\bB$:
$$
A_{\bf q,k}=\alphab^T(\bq )\etab(\bk ),
$$
where
\begin{eqnarray}
\etab=
\left(
\matrix{\etab_1\cr\etab_2}
\right)=
\left(
\matrix{
\eta_{{\bf l}_1,1}\cr
\vdots\cr
\eta_{{\bf l}_{z/2},1}\cr
\eta_{{\bf l}_1,2}\cr
\vdots\cr
\eta_{{\bf l}_{z/2},2}
}
\right),&&
\eta_{{\bf l},1}(\bk )=
\cos({\bf kl})
\ \ \
\eta_{{\bf l},2}(\bk )=
\sin({\bf kl}).
\end{eqnarray}
Thus we can rewrite the expression for the function $\Sigma$ as follows
\begin{eqnarray}
\Sigma(\cE,\bq )&=&[\bB({\bf 1+B})^{-1}\bA]_{\bf q,q}=\nonumber\\
&&\alphab^T(\bq )
(1+\gammab(\cE))^{-1}\gammab(\cE))
\etab(\bq )=
\alphab^T(\bq )
\etab(\bq )-
\alphab^T(\bq )
(1+\gammab(\cE))^{-1}
\etab(\bq ).
\end{eqnarray}

The calculation of the inverse matrixe $(1+\gammab(\cE))^{-1}$ can be easily
carried out for any simple lattice. Let us consider as an example the
$d$--dimensional simple cubic lattice. Then we have
$$
({\bf kl})=k_i a,\ \ \  i=1,\ldots,d,
$$
where $a$ is a lattice constant. The matrix $\gammab$ takes the form
\begin{eqnarray}
\gammab=
\left(
\matrix
{
\matrix{
 \phi   & \chi          & \ldots        & \chi     \cr
 \chi   & \ddots        & \ddots        & \vdots   \cr
 \vdots & \ddots        & \ddots        & \chi     \cr
 \chi   & \ldots        & \chi          & \phi
        }
&{\bf 0}\cr
{\bf 0}&
\matrix{
 \psi   & 0         & \ldots             & 0        \cr
 0      & \ddots    & \ddots             & \vdots   \cr
 \vdots & \ddots    & \ddots             & 0        \cr
 0      & \ldots    & 0                  & \psi
        }
\cr
}
\right).
\end{eqnarray}
All the diagonal elements of the matrix $\gammab_1$ read
\begin{equation}
\phi={1\over d}{1\over N}\sum_\bk 
{\cos^2(k_1a)-\cos(k_1a)
\over
\cE-\cE_0^{\rm sc}(\bk )
}={1\over d}\left(
I_{c^2}^d(\cE)-I_{c}^d(\cE)
\right),
\end{equation}
where as the off--diagonal elements read
\begin{equation}
\chi={1\over d}{1\over N}\sum_\bk 
{\cos(k_1a)\cos(k_2a)-\cos(k_1a)
\over
\cE-\cE_0^{\rm sc}
}={1\over d}\left(
I_{cc}^d(\cE)-I_{c}^d(\cE)
\right).
\end{equation}
The matrix $\gammab_2=\psi\hat{1}$, where
\begin{equation}
\psi={1\over d}{1\over N}\sum_\bk 
{\sin^2(k_1a)
\over
\cE-\cE_0^{\rm sc}(\bk )
}={1\over d}
I_{s^2}^d(\cE).
\end{equation}
Here
\begin{equation}
\cE_0^{\rm sc}(\bk )=
1-{1\over d}\sum\limits_{i=1}^d\cos(k_ia)
\end{equation}
is a dimensionless energy of spin excitations
of the non--dilute simple cubic ferromagnet and
we have introduced the following notation
\begin{eqnarray}
I_{c^2}^d(\cE)={1\over N}\sum_\bk 
{\cos^2(k_1a)
\over
\cE-\cE_0^{\rm sc}(\bk )
},&&
I_{c}^d(\cE)={1\over N}\sum_\bk 
{\cos(k_1a)
\over
\cE-
\cE_0^{\rm sc}(\bk )
},\\ \nonumber
I_{cc}^d(\cE)={1\over N}\sum_\bk 
{\cos(k_1a)\cos(k_2a)
\over
\cE-
\cE_0^{\rm sc}(\bk )
},&&
I_{s^2}^d(\cE)={1\over N}\sum_\bk 
{\sin^2(k_1a)
\over
\cE-
\cE_0^{\rm sc}(\bk )
}.
\end{eqnarray}

Let us note that the sum $\sum\limits_\bk\betab$ that appeared in the expressions
for the functions $C$ can be rewritten as
\begin{eqnarray}
\sum\limits_\bk\betab_1(\cE,\bk )=I_{c}^d(\cE)
\left(
\matrix
{
1\cr\vdots\cr 1
}
\right),&&
\sum\limits_\bk\betab_2(\cE,\bk )={\bf 0}.
\end{eqnarray}
It can be easily verified directly that
\begin{eqnarray}
(1+\gammab_1)^{-1}=
{1/d
\over
1+\phi+(d-1)\chi}
\left(
\matrix{
1       &\ldots &1      \cr
\vdots  &\ddots &\vdots \cr
1       &\ldots &1
}
\right)+
{1/d
\over
1+\phi-\chi}
\left(
\matrix{
d-1     &-1     &\ldots &-1     \cr
-1      &\ddots &\ddots &\vdots \cr
\vdots  &\ddots &\ddots &-1     \cr
-1      &\ldots &-1     &d-1
}
\right)
\end{eqnarray}
Now we can rewrite the expressions for the functions $C$ as follows
\begin{eqnarray}
C(\cE,\bq )&=&
\alphab_1^T(\bq )
(1+\gammab_1(\cE))^{-1}
\betac_1(\cE,\bk )= 
%\\\nonumber&&
{
-I_{c}^d(\cE)\cE_0^{\rm sc}(\bq )
\over
1+\phi+(d-1)\chi
}=
{
\cE_0^{\rm sc}(\bq )
\over
\cE-{1\over I_{c}^d(\cE)}
},
\end{eqnarray}
\begin{eqnarray}
C_a(\cE,\bq )&=&
\alphab_1^T(\bq )
\left[1+
2(1+2\gammab_1(\cE))^{-1}
-(1+\gammab_1(\cE))^{-1}
\right]
\betac_1(\cE)=\\\nonumber
&&-I_{c}^d(\cE)\cE_0^{\rm sc}(\bq )-{2I_{c}^d(\cE)\cE_0^{\rm sc}(\bq )
\over
1+2\phi+2(d-1)\chi
}
+
{I_{c}^d(\cE)\cE_0^{\rm sc}(\bq )
\over
1+\phi+(d-1)\chi
}
=\\\nonumber&&
\cE_0^{\rm sc}(\bq )
\left[
-I_{c}^d(\cE)+
{
1
\over
\cE-{1\over 2 I_{c}^d(\cE)}
}-
{
1
\over
\cE-{1\over I_{c}^d(\cE)}
}
\right],
\end{eqnarray}
\begin{eqnarray}
C_b(\cE,\bq )=
\alphab_1^T(\bq )
\sum\limits_\bk\betab_1(\cE,\bk )
%\\\nonumber=&&
-I_{c}^d(\cE)\cE_0^{\rm sc}(\bq ),
\end{eqnarray}
where we have used the fact that
\begin{eqnarray}
0&=&{1\over N}\sum_\bk{1\over d}\sum_i\cos(k_i a)=
{1\over N}\sum_\bk{1\over d}\sum_i\cos(k_i a)
{{\cal E}-1+{1\over d}\sum_j\cos(k_j a)
\over
\cE-1+{1\over d}\sum_l\cos(k_l a)}=\nonumber\\
&&(\cE-1)I_{c}^d(\cE)+{1\over d}I_{c^2}^d(\cE)+{d-1\over d}I_{cc}^d(\cE)
\end{eqnarray}
and
\begin{eqnarray}
\phi+(d-1)\chi={1\over d}I_{c^2}^d(\cE)+{d-1\over d}I_{cc}^d(\cE)
-I_{c}^d(\cE)=-\cE I_{c}^d(\cE)
\end{eqnarray}

We can obtain the expression for the function $\Sigma$ in a similar way.
\begin{eqnarray}
\Sigma(\cE,\bq )\nonumber
&&=\alphab^T(\bq )
(1+\gammab(\cE))^{-1}\gammab(\cE))
\etab(\bq )=\\
&&\alphab_1^T(\bq )
\etab_1(\bq )-
\alphab_1^T(\bq )
(1+\gammab(\cE))^{-1}
\etab_1(\bq )+{\psi\over 1+\psi}\alphab_2^T(\bq )
\etab_2(\bq )=\nonumber\\
&&{1\over d}\sum_i(\cos(k_i a)-1)\cos(k_i a)+
{I_{s^2}^d(\cE)
\over
d+I_{s^2}^d(\cE)
}
{1\over d}\sum_i\sin^2(k_i a)-\nonumber\\
&&
{1\over 1-\cE I_{c}^d(\cE)}
{1\over d^2}\sum_{i,j}(\cos(k_i a)-1)\cos(k_j a)-
\nonumber\\
&&{1\over d+I_{c^2}^d(\cE)-I_{cc}^d(\cE)}
\left(\sum_i(\cos(k_i a)-1)\cos(k_i a)-{1\over d}\sum_{i,j}(\cos(k_i a)-1)\cos(k_j a)
\right)=\nonumber\\
&&\cE_0(\bk )-2\cE_1(\bk )+2{I_{s^2}^d(\cE)
\over
d+I_{s^2}^d(\cE)
}\cE_1(\bk )+\nonumber\\
&&{1\over 1-\cE I_{c}^d(\cE)}
(\cE_0(\bk )-\cE_0^2(\bk ))+
{d\over d+I_{c^2}^d(\cE)-I_{cc}^d(\cE)}
(2\cE_1(\bk )-2\cE_0(\bk )+\cE_0^2(\bk )),
\end{eqnarray}
where
\begin{equation}
\cE_1(\bk )={1\over 2d}\sum_i\sin^2(k_i a).
\end{equation}

Let us consider the long--wave solutions of equation (\ref{Spectrum})
for the spin excitation spectrum.
Taking into account that
\begin{eqnarray}
\cE_0(\bk )={a^2\over 2d}k^2+o(k^2)&{\rm and}&
\cE_1(\bk )={a^2\over 2d}k^2+o(k^2)
\end{eqnarray}
we can easily find that
\begin{eqnarray}
\Sigma(\cE,\bq )=-2
{a^2\over 2d}
{
I_{s^2}(d)
\over 1-I_{s^2}(d)
}k^2
+o(k^2),
\end{eqnarray}
where
\begin{eqnarray}
I_{s^2}(d)=-{1\over d}I_{s^2}^d(0)={1\over N}\sum_\bk{\sin^2(k_1a)
\over\sum_i(1-\cos(k_ia))}.
\end{eqnarray}

\setcounter{equation}{0}
\section{Calculation of the Spectral Density $g_{2a}$ 
for the Simple Cubic Lattice}
Let us calculate the function 
\begin{eqnarray}\label{ga_ad}
g_{2a}(\cE)\equiv-{1\over 2\pi N}{\rm Sp}\Im\overline{\Gsad}
\end{eqnarray}
we need to obtain the low--temperature behaviour of the structural 
quadratic fluctuations of magnetization $\ds$. Using the result of the 
previous section we get the $d$--dimensional ($sc$) lattice
\begin{eqnarray}\label{ga_ap}
g_{2a}(\cE)&\equiv&-{1\over 2\pi N}{\rm Sp}\Im\overline{\Gsad}=
-{1\over \pi N}\Im\sum_\bq 
\left.
{
C_a(\cE,\bq )
\over
\cE-\cE_0(\bq )
}
\right|_{{\cal E}+i0}=\\\nonumber&&
-{1\over \pi N}\Im\sum_\bq 
\left.
{
\cE_0^{\rm sc}(\bq )
\over
\cE-\cE_0^{\rm sc}(\bq )
}\left[
-I_{c}^d(\cE)+
{
1
\over
\cE-{1\over 2 I_{c}^d(\cE)}
}-
{
1
\over
\cE-{1\over I_{c}^d(\cE)}
}
\right]
\right|_{{\cal E}+i0}=\\\nonumber&&
{1\over \pi}\Im
\left.
U_\cE
\left[
-I_{c}^d(\cE)+
{
1
\over
\cE-{1\over 2}V_\cE
}-
{
1
\over
\cE-V_\cE
}
\right]
\right|_{{\cal E}+i0}.
\end{eqnarray}
Here 
\begin{eqnarray}
U_\cE=
-{1\over N}\sum_\bq 
{
\cE_0^{\rm sc}(\bq )
\over
\cE-\cE_0^{\rm sc}(\bq )
}=
I_{c}^d(\cE)-I_{1}^d(\cE),\ \ \
I_1^d(\cE)={1\over N}\sum_\bk 
{1
\over
\cE-
\cE_0^{\rm sc}(\bk )
},\ \ \ 
V_\cE={1\over I_{c}^d(\cE)}.
\end{eqnarray}
Simple calculations give us the following expression for the spectral density
\begin{eqnarray}
g_{2a}(\cE)=
{1\over \pi}
\left\{ 
	U'_\cE
	\left[	{I_{c}^d}''(\cE)+
		{
		2V''_\cE 
		\over
		(2\cE-V'_\cE)^2+(V''_\cE)^2
		}-
		{
		V''_\cE
		\over
		(\cE-V'_\cE)^2+(V''_\cE)^2
		}
	\right]-
\right.\\\nonumber
\left. 
	U''_\cE
	\left[	-{I_{c}^d}'(\cE)+
		{
		4\cE-2V'_\cE
		\over
		(2\cE-V'_\cE)^2+(V''_\cE)^2
		}-
		{
		\cE-V'_\cE
		\over
		(\cE-V'_\cE)^2+(V''_\cE)^2
		}
	\right]
\right\},
\end{eqnarray}
where 
\begin{eqnarray}
U_{{\cal E}+i0}=U'_\cE-iU''_\cE,
%&&
\ \ \ 
V_\cE={1\over I_{c}^d(\cE)},
%\\\nonumber
\ \ \ 
V_{{\cal E}+i0}={1\over {I_{c}^d}'(\cE)-i{I_{c}^d}''(\cE)}=
V'_\cE+iV''_\cE.\end{eqnarray}

Let us consider an asymptotic behaviour of the spectral density 
$g_{2a}(\cE)$ at small $\cal E$.
\begin{eqnarray}
&&U'_\cE=-{{\cal P}\over N}\sum_\bk 
{
\cE_0^{\rm sc}(\bk )
\over
\cE-\cE_0^{\rm sc}(\bk )
}=1-\cE
{{\cal P}\over N}\sum_\bk 
{
1
\over
\cE-\cE_0^{\rm sc}(\bk )
}
\mathop{\to}\limits_{{\cal E}\to 0}1,\\
&&U''_\cE=-{\pi\over N}\sum_\bk 
\cE_0^{\rm sc}(\bk )
\deltaF
\left(
\cE-\cE_0^{\rm sc}(\bk )
\right)
=-\pi\cE
g_0(\cE)
,\\
&&{I_{c}^d}'(\cE)=
{{\cal P}\over N}\sum_\bk 
{
1-\cE_0^{\rm sc}(\bk )
\over
\cE-\cE_0^{\rm sc}(\bk )
}
\\
&&{I_{c}^d}''(\cE)=
{\pi\over N}\sum_\bk 
\left(
1-\cE_0^{\rm sc}(\bk )
\right)
\deltaF
\left(
\cE-\cE_0^{\rm sc}(\bk )
\right)=
\pi(1-\cE)
g_0(\cE)
\mathop{\to}\limits_{{\cal E}\to 0}
\pi g_0(\cE),\\
&&V'_\cE=
{{I_{c}^d}'(\cE)
\over
\left({I_{c}^d}'(\cE)\right)^2
+\left({I_{c}^d}''(\cE)\right)^2
},\ \ \
V''_\cE=
{{I_{c}^d}''(\cE)
\over
\left({I_{c}^d}'(\cE)\right)^2
+\left({I_{c}^d}''(\cE)\right)^2
}.
\end{eqnarray}
Here
\begin{equation}
g_0(\cE)={1\over N}\sum_\bk 
\deltaF
\left(
\cE-\cE_0^{\rm sc}(\bk )
\right)
\end{equation}
is the density of states of a non--dilute crystal.
Finally we get 
\begin{eqnarray}
g_{2a}(\cE)
\mathop{\to}\limits_{{\cal E}\to 0}
{1\over \pi}
\left(
U''_\cE{I_{c}^d}'(\cE)+{I_{c}^d}''(\cE)U'_\cE+
{
U''_\cE V'_\cE+V''_\cE U'_\cE
\over
(V'_\cE)^2+(V''_\cE)^2
}
\right)
=
{2\over \pi}
\left(
U''_\cE{I_{c}^d}'(\cE)+{I_{c}^d}''(\cE)U'_\cE
\right)
\mathop{\to}\limits_{{\cal E}\to 0}
2g_0(\cE)
.
\end{eqnarray}
\end{appendix}

\end{document}